\documentclass[pra,aps,10pt,preprint,showkeys,preprintnumbers,superscriptaddress,floatfix]{revtex4-1}

\usepackage{graphicx}		
\usepackage{bm}				
\usepackage{amsmath}
\usepackage{color}

\renewcommand{\deg}{$^\circ$}

\begin{document}

\title{Excitation and propagation of spin waves in non-uniformly magnetized waveguides}

\author{Frederic Vanderveken}
\affiliation{Imec, 3001 Leuven, Belgium}
\affiliation{KU Leuven, Departement Materiaalkunde, SIEM, 3001 Leuven, Belgium}
\author{Hasnain Ahmad}
\affiliation{Imec, 3001 Leuven, Belgium}
\author{Marc Heyns}
\affiliation{Imec, 3001 Leuven, Belgium}
\affiliation{KU Leuven, Departement Materiaalkunde, SIEM, 3001 Leuven, Belgium}
\author{Bart Sor\'ee}
\affiliation{Imec, 3001 Leuven, Belgium}
\affiliation{KU Leuven, Departement Elektrotechniek, TELEMIC, 3001 Leuven, Belgium}
\affiliation{Universiteit Antwerpen, Departement Fysica, 2000 Antwerpen, Belgium}
\author{Christoph Adelmann}
\affiliation{Imec, 3001 Leuven, Belgium}
\author{Florin Ciubotaru}
\email[Author to whom correspondence should be addressed. Electronic mail: ]{\texttt{Frederic.Vanderveken@imec.be}, \texttt{Florin.Ciubotaru@imec.be}} 
\affiliation{Imec, 3001 Leuven, Belgium}

\date{\today}

\begin{abstract}
The characteristics of spin waves in ferromagnetic waveguides with nonuniform magnetization have been investigated for situations where the shape anisotropy field of the waveguide is comparable to the external bias field. Spin-wave generation was realized by the magnetoelastic effect by applying normal and shear strain components, as well as by the Oersted field emitted by an inductive antenna. The magnetoelastic excitation field has a nonuniform profile over the width of the waveguide because of the nonuniform magnetization orientation, whereas the Oersted field remains uniform. Using micromagnetic simulations, we indicate that both types of excitation fields generate quantised width modes with both odd and even mode numbers as well as tilted phase fronts. We demonstrate that these effects originate from the average magnetization orientation with respect to the main axes of the magnetic waveguide. Furthermore, it is indicated that the excitation efficiency of the second-order mode generally surpasses that of the first-order mode due to their symmetry. The relative intensity of the excited modes can be controlled by the strain state as well as by tuning the dimensions of the excitation area. Finally, we demonstrate that the nonreciprocity of spin-wave radiation due to the chirality of an Oersted field generated by an inductive antenna is absent for magnetoelastic spin-wave excitation.
\end{abstract}


\maketitle

Magnonic devices that use spin waves as information carriers are currently intensely researched due to their great potential for ultralow power computing technologies. Such technologies can be envisaged to complement the predominant complementary metal-oxide-semiconductor (CMOS) circuits in future technology nodes \cite{schneider08,chumak17, chumak15, wang19}. In magnonic devices, the information can be encoded in the amplitude and/or the phase of the spin wave and is transported along magnetic waveguides. When using phase coding, spin wave interference can be used to design magnonic majority gates \cite{khitun1,fischer17,Talmelli19} that can be advantageous for logic circuit design \cite{Radu_iedm15}. Thus, future hybrid spin wave–-CMOS technology promises to operate at much lower power and occupies less area per computing throughput with respect to conventional CMOS.

In competitive hybrid spin wave--CMOS technology, the density of logic functions in the circuit must be high and therefore the magnonic devices must be miniaturized to the nanoscale. Yet, scaling magnetic waveguides leads to higher aspect ratios and introduces stronger shape anisotropy. Typical spin-wave devices are commonly operated in the Damon-Eshbach geometry with a magnetic bias field applied transversely to the waveguide. For ``macroscopic'' waveguides, the shape anisotropy is negligible, and the magnetization is uniform and parallel to the applied external bias field. In contrast, for waveguide widths of the order of (a few) 10 nm, the shape anisotropy field becomes much stronger than typical values of the transverse bias field. In this case, the magnetization is uniformly aligned along the waveguide direction. 

Between these two regimes, a ``mesoscopic'' regime exists where the magnitude of the shape anisotropy field is comparable to typical external bias fields. Furthermore, the shape anisotropy field amplitude is position dependent and typically stronger near the edges of the waveguide. In this case, the total effective magnetic field--and thus the magnetization---is tilted with respect to the principal axes of the waveguide \cite{jorzick02,demidov15}. This magnetization tilt is the highest near the edges of the waveguide as compared to the middle. As a result, the magnetization is nonuniformy oblique oriented inside the waveguide. Recently, this regime has been studied experimentally with increasing interest as a starting point for future nanoscale magnonic devices \cite{ciubotaru1,Meso1,Meso2}. It is clear that the interpretation of such experiments as well as the design and optimization of magnonic devices at the mesoscale requires a thorough understanding of how the tilt and nonuniformity of the magnetization affect the generation and propagation of spin waves.

A key parameter of hybrid spin wave–-CMOS circuits is the efficiency of the transduction mechanism between electronic and magnonic subsystems. Different spin-wave excitation mechanisms have been proposed including excitation by inductive antennas \cite{chumak09,demidov09,pirro14, ciubotaru1}, spin transfer torques \cite{madami11}, or spin orbit torques \cite{divinskiy18,talmelli1}. Microwave antennas utilise electric currents to generate spin waves. In contrast, spin transfer and spin orbit torques depend on the current density and are thus better scalable. However, spin-wave generation by these mechanisms is not energy efficient. Recently, magnetoelectric transducers have been proposed as an alternative route to excite spin waves with potentially low energy consumption \cite{khitun1,weiler11,cherepov14,Radu_iedm15,duflou17}. Magnetoelectric transducers combine piezoelectric and magnetoelastic effects to excite spin waves by application of voltages (electric fields) instead of currents, leading to much better scalability. Although both experiments \cite{cherepov14} and micromagnetic simulations \cite{duflou17,chen17} have demonstrated the possibility to excite spin waves by magnetoelectric effects, a detailed understanding of the underlying magnetoelastic phenomena in scaled waveguides is still lacking. 

In this paper, we investigate by means of micromagnetic simulations the propagation characteristics of spin waves in nonuniformly magnetized mesoscale waveguides. We further analyse the excitation of these spin waves by nonuniform magnetoelastic excitation fields and compare it with the uniform Oersted excitation field.  The comparison allows us to separate between effects of the nonuniform excitation field and of the nonuniform tilted magnetization itself. In addition, two different strain states are applied to analyze how different dynamic strain components couple to the magnetization.

In the following, first, the simulation model is introduced. Then, the spin-wave modes in a nanoscale waveguide are discussed, their dispersion relations are presented, and the observed noncolinearity of phase and group velocities is explained. Subsequently, the relative excitation intensity of the different modes is investigated for and related to geometries of the excitation fields. Finally, the relative excitation efficiency of the nonuniform magnetoelastic and uniform Oersted field and the nonreciprocity of spin-wave emission is discussed. 

The studied structures were based on CoFeB waveguides with a length of 10 $\mu$m, a width of 200 nm, and a thickness of 10 nm. Typical magnetic parameters for CoFeB were used: a saturation magnetization of $M_s = 1.25 \times 10^6 $A/m \cite{conca13}, an exchange stiffness constant of $A_{ex}=1.89 \times 10^{-11}$ J/m$^3$ \cite{conca14}, a gyromagnetic ratio of $g = 2$, and a Gilbert damping of $\alpha$ = 0.004. To avoid spin-wave reflection, the damping constant was smoothly increased to $\alpha = 0.2$ within 1 $\mu$m from the waveguide ends. Experimental CoFeB waveguides are typically amorphous or polycrystalline and thus the magnetocrystalline anisotropy was assumed to be zero.

\begin{figure}
\centering
\includegraphics[width=8.2 cm]{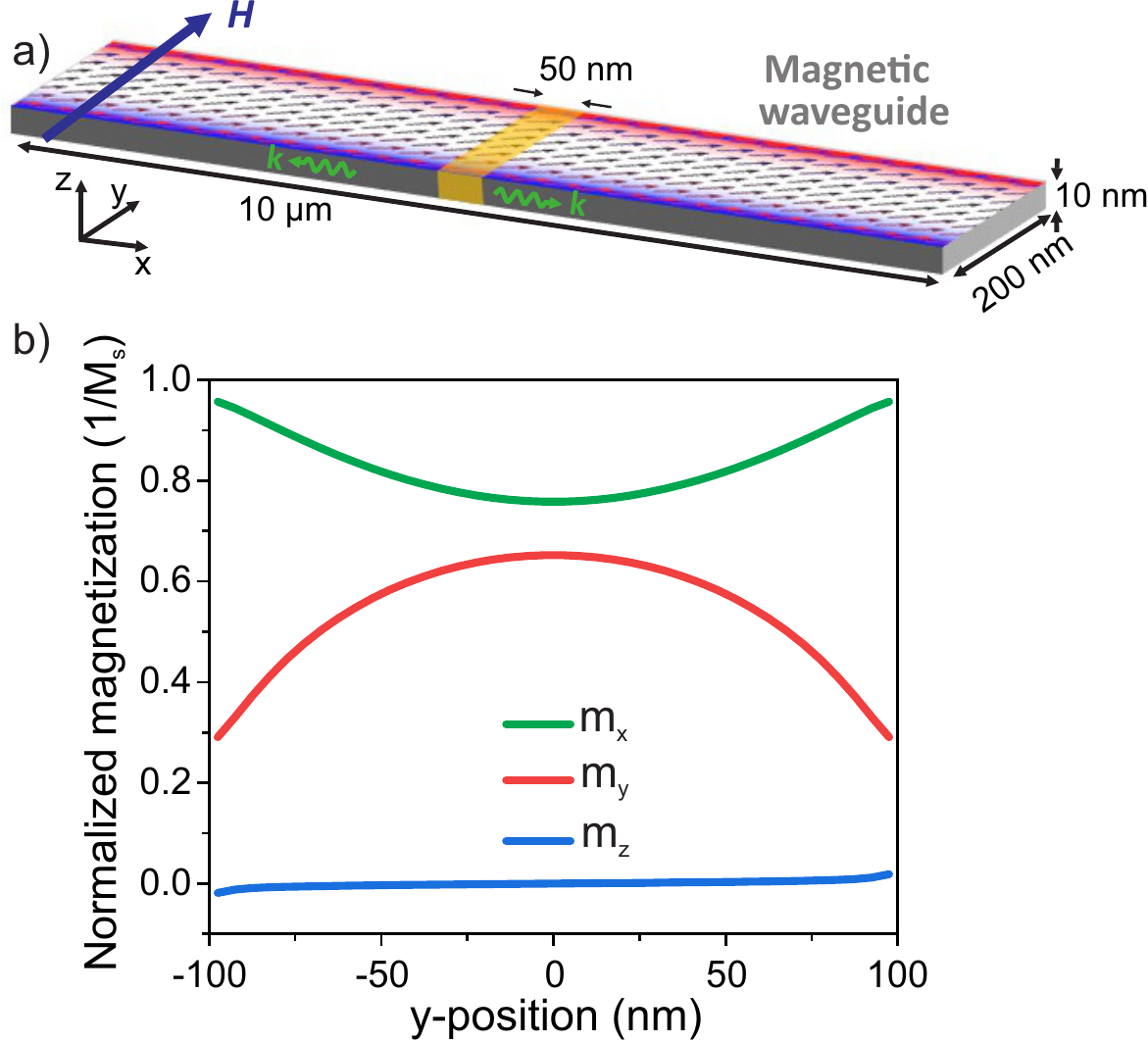}
\caption{\textbf{(a)} Sketch of the waveguide geometry. The arrows represent the direction of the in-plane equilibrium magnetization whereas the out-of-plane component of the magnetization is represented by the color map. The yellow region in the middle represents the excitation region where the excitation field is applied. \textbf{(b)} Magnetization components $m_x$, $m_y$ and $m_z$ in the equilibrium ground state across the waveguide. The external magnetic bias field was $\mu_0 H_\mathrm{ext} = 50$ mT.}
\label{fig:geo}
\end{figure}

An external magnetic bias field was applied in-plane in the $y$-direction, \emph{i.e.}~transverse to the waveguide. In this way, the bias field counteracts the shape anisotropy, which tries to align the magnetization in the $x$-direction along the waveguide. Except when stated otherwise, the amplitude of the bias field was set to $\mu_0 H_\mathrm{ext} = 50$ mT (a magnitude commonly used in experiments), which resulted in a magnetization state that was not completely saturated along the bias field direction. Moreover, the magnetization orientation varies over the width of the waveguide because of the nonuniformity of the demagnetization field. A sketches of the geometry, magnetic fields and reference system is graphically depicted in Fig.~\ref{fig:geo}a. The magnetization components along the $x$-, $y-$ and $z$-directions in the waveguide are indicated in Fig.~\ref{fig:geo}b, which clearly establishes the orientation and the nonuniformity of the magnetization.

Two spin-wave excitation mechanisms were considered: (i) oscillating Oersted fields and (ii) oscillatory magnetoelastic fields. The Oersted field is typically generated via an inductive antenna placed on top of the magnetic waveguide. The AC current inside the antenna wire generates an oscillating Oersted field inside the magnetic waveguide, which generates spin waves. The amplitude and shape of the Oersted field only depends on the shape of and current in the antenna wire. Hence, this excitation field is independent of the magnetization orientation inside the magnetic waveguide. 

The second approach for the excitation of spin waves is based on magnetoelastic coupling, which describes the interaction between the elastic and magnetic domain. If strain is present inside the magnetostrictive material, then this strain affects the magnetization orientation, i.e. inverse magnetostriction or the Villari effect. This coupling can be represented by a magnetoelastic effective field $\bm{H}_{mel}$, which depends on both the strain tensor $\bm{\varepsilon}$ and the normalized magnetization vector $\bm{m} = \left( m_x, m_y, m_z\right)$ and is given by \cite{kitte49}

\begin{equation}
\label{eq:mel-field}
\bm{H}_{mel} = -\frac{2}{\mu_0M_s}  \begin{pmatrix}
B_1\varepsilon_{xx}m_x + B_2(\varepsilon_{xy}m_y+\varepsilon_{zx}m_z) \\
B_1\varepsilon_{yy}m_y + B_2(\varepsilon_{xy}m_x+\varepsilon_{yz}m_z) \\
B_1\varepsilon_{zz}m_z + B_2(\varepsilon_{zx}m_x+\varepsilon_{yz}m_y) 
\end{pmatrix}.
\end{equation}

\noindent Here, $B_1$ and $B_2$ are the magnetoelastic coupling constants of the waveguide material. Hence, in contrast with the Oersted excitation field, the magnetoelastic excitation field depends on the magnetization orientation. Therefore, nonuniform magnetization orientation inside the waveguide also results in nonuniform magnetoelastic excitation fields. Spin waves excited by the magnetoelastic fields are thus potentially affected not only by the nonuniformity of the magnetization in the waveguide itself but also by the nonuniformity of the excitation field.

As mentioned in the introduction, this magnetoelastic coupling is a crucial part for magnetoelectric transducers that consist of magnetostrictive and piezoelectric materials. For spin wave excitation, these transducers are locally attached to the magnetic waveguide. Hence, by applying a voltage to the transducer, a strain is induced inside the magnetic waveguide and the magnetoelastic excitation field is created.

The resulting strain state inside the waveguide can become very complex and is nontrivial at GHz frequencies and at nanoscale dimensions. Furthermore, the strain state is strongly affected by numerous parameters such as the dimensions of the transducer, material choice of the transducer, dimensions of the waveguide, and effects such as the substrate clamping and different edge effects.  The description and behavior of this complex strain states is a pure elastodynamic problem and not the goal of this paper. Furthermore, these complex strain states result in complex magnetoelastic excitation fields which does not allow to gain qualitative understanding of the coupling. Therefore, here, two simplistic strain states are assumed which allow to give more insight into the coupling of the different strain components with the nonuniform magnetization. This insight can then be used to explain more complex and realistic excitation strain fields. 

The two different strain states used in this paper are based on the work of \cite{duflou17}. The first strain state can be considered as a simple expansion in the y-direction of a cuboid and contraction in the two other directions opposite to that direction. In this case, only the three normal strain components are different from zero and their value is determined by the dimensions of the magnetic waveguide and the transducer. Therefore, this state is named the normal strain state $\bm{\varepsilon}_n$. The second strain state utilised in this paper is obtained by rotating the previous state in-plane by 45\deg{}. This rotation makes the in-plane normal strains $\varepsilon_{pl}$ equal to each other and introduces the shear strain element $\varepsilon_{xy}$. Therefore, this state is named the shear strain state $\bm{\varepsilon}_s$. Both strain tensors are given as follows

\begin{equation}
\label{eq:strain-tensor0}
\bm{\epsilon}_n = \begin{pmatrix}
-\epsilon_{xx} & 0 & 0 \\
0 & \epsilon_{yy} & 0 \\
0 & 0 & -\epsilon_{zz}
\end{pmatrix}; 
\hspace{20 pt}
\bm{\epsilon}_{s} = \bm{R}(\theta)^\mathrm{T} \bm{\varepsilon}_b \bm{R}(\theta) = \begin{pmatrix}
\epsilon_{pl} & \epsilon_{xy} & 0 \\
\epsilon_{xy} & \epsilon_{pl} & 0 \\
0 & 0 & -\epsilon_{zz}
\end{pmatrix},
\end{equation}

\noindent with $\bm{R}(\theta)$ the rotation matrix. Furthermore, in the general case, these strain components can also be space dependent. 

These strain states can be simplified by adding or subtracting hydrostatic strain tensors, in which the diagonal components are all equal and the off-diagonal components are zero, \emph{i.e.} $\epsilon_{xx}=\epsilon_{yy}=\epsilon_{zz}$ and $\epsilon_{xy}=\epsilon_{xz}=\epsilon_{yz}=0$. The general equation of the torque exerted by an arbitrary strain state on the magnetization is given by
\begin{equation}
\begin{split}
	\mathbf{T} = & \mu_0 \mathbf{M}\times \mathbf{H}_\mathrm{mel} \\
			& = \frac{2}{M_s^2}\begin{bmatrix}
	B_1 m_ym_z(\epsilon_{yy}-\epsilon_{zz})+B_2\left(m_xm_z\epsilon_{xy}-m_xm_y\epsilon_{zx} +\left( m_z^2-m_y^2 \right)\epsilon_{yz}\right) \\
	B_1 m_xm_z(\epsilon_{zz}-\epsilon_{xx})+B_2\left(m_xm_y\epsilon_{yz}-m_zm_y\epsilon_{xy} +\left( m_x^2-m_z^2 \right)\epsilon_{zx}\right) \\
	B_1 m_xm_y(\epsilon_{xx}-\epsilon_{yy})+B_2\left(m_ym_z\epsilon_{zx}-m_zm_x\epsilon_{yz} +\left( m_y^2-m_x^2 \right)\epsilon_{xy}\right)
	\end{bmatrix} .
\end{split}
\end{equation}
\noindent It is clear from this equation that hydrostatic strain states do not exert any torque on the magnetization and thus have no influence on the magnetization dynamics \cite{duflou17}. Since the torque is linear in the strain tensor components, suitable hydrostatic strain states can be added (or subtracted) to obtain simplified, magnetically equivalent strain tensors. If $\varepsilon_{zz}$ is added to all the diagonal elements of the normal strain case and $\varepsilon_{pl}$ is subtracted from all the diagonal elements of the shear case, one obtains the two following magnetically equivalent normal and shear strain states

\begin{equation}
\label{eq:strain-tensor}
\bm{\varepsilon}_n = \begin{pmatrix}
-\varepsilon'_{xx} & 0 & 0 \\
0 & \varepsilon'_{yy} & 0 \\
0 & 0 & 0
\end{pmatrix}; 
\hspace{20 pt}
\bm{\varepsilon}_{s} = \begin{pmatrix}
0 & \varepsilon_{xy} & 0 \\
\varepsilon_{xy} & 0 & 0 \\
0 & 0 & -\varepsilon'_{zz}
\end{pmatrix}.
\end{equation}

Hence, there are only two independent strain components which describe the coupling. Furthermore, it is clear that the normal strain state is ideal to study the coupling of the in-plane normal strains with the magnetization. On the other hand, the shear state is ideal to study the coupling between the in-plane shear strain and the magnetization. Note that the z-component in the shear strain case has negligible effect because the z-component of the equilibrium magnetization is also very weak. 

The excitation fields were locally applied in the middle of the waveguide over a width of 50 nm as indicated in Fig.~\ref{fig:geo}a. For the Oersted field originating from an antenna, this is a good approximation if the antenna width is larger than the antenna thickness. For the magnetoelastic excitation field originating from the magnetoelectric transducer, this is a good approximation if the transducer has a high Q-factor, i.e. there is very little elastodynamics leakage outside the transducer.

The magnetization dynamics were simulated using the object-oriented micromagnetic framework (OOMMF) \cite{OOMMF}. The YY\_mel module was utilised to compute the magnetoelastic field \cite{yahagi14} based on the strain tensors of equation \ref{eq:strain-tensor}. For the CoFeB material parameters in the previous section, the exchange length is 4.4 nm. Therefore, the mesh size was set to $5 \times 5 \times 5$ nm$^3$, which is comparable to the exchange length, as typically done in micromagnetic simulations. A uniform strain distribution was assumed in the excitation region and isotropic magnetoelastic coupling constants $B_1=B_2=8.85$~MJ/m$^3$ were considered, which are typical for polycrystalline CoFeB films \cite{gueye16}.

\begin{figure*}
\centering
\includegraphics[width=16cm]{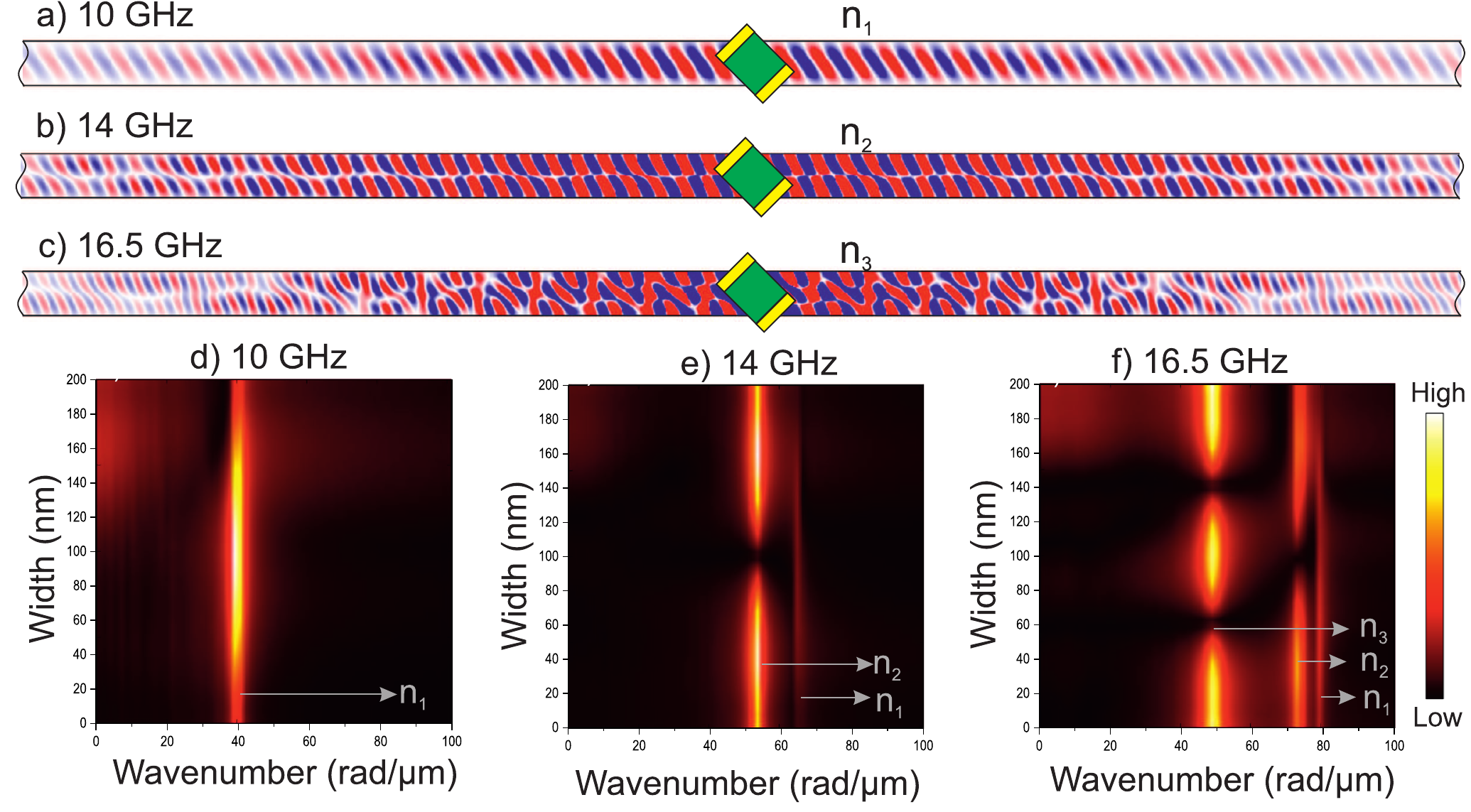}
\caption{\textbf{(a--c)} Snapshot images of the $z$-component of the magnetization dynamics excited by the magnetoelastic shear strain state in a 200-nm-wide CoFeB waveguide at different frequencies after 7 ns, \emph{i.e.}~under steady-state conditions. \textbf{(d--f)} Spin-wave wavenumber distribution across the waveguide obtained by a spatial Fourier transform of the $z$-component of the magnetization dynamics under steady-state conditions for magnetoelastic shear excitation at the three excitation frequencies of 10, 14, and 16.5 GHz, respectively. The external transverse magnetic bias field was $\mu_0 H_\mathrm{ext} = 50$ mT.}
\label{fig:modes}
\end{figure*}

In a first step, the excitation and propagation characteristics of spin waves in nanoscale waveguides with a transverse magnetic bias field were studied at different excitation frequencies. As an example, Figs.~\ref{fig:modes}a--c indicate snapshot images of the steady state magnetization dynamics generated by the magnetoelastic field from the shear strain state at frequencies of 10, 14, and 16.5~GHz, respectively. The simulations demonstrate the generation of propagating spin waves, albeit with a tilted phase front with respect to the waveguide direction. In addition, the dynamic magnetization profiles at the different frequencies indicate the excitation of quantised spin-wave modes.

To identify the excited spin-wave modes at each frequency, a spatial Fourier transform was performed along the length of the waveguide ($x$-direction). The results are indicated in Figs.~\ref{fig:modes}d--f as a function of the spin-wave wavenumber and the $y$-position across the waveguide. At a frequency of 10 GHz, only a first-order mode, labeled $n_1$, was excited (Fig.~\ref{fig:modes}d). In contrast, at a frequency of 14 GHz, both first- ($n_1$) and second-order ($n_2$) modes were observed (Fig.~\ref{fig:modes}e). It is worth noting that the intensity of the second-order mode excited at 14 GHz was much higher than the intensity of the first-order mode. This is in strong contrast to the case of spin-wave excitation by Oersted fields from an inductive antenna in the ideal Damon-Eshbach or backward-volume geometries, where the excitation of the second-order mode is forbidden by symmetry. Finally, at a frequency of 16.5 GHz, a third-order mode, labeled $n_3$, appeared (Fig.~\ref{fig:modes}f). 

Figure~\ref{fig:modes} also indicates that the spin-wave intensity did not fall to zero at both edges of the waveguide. This is attributed to dipolar pinning near the edges and can be described by an effective width of the waveguide, which is larger than its physical width \cite{wang18}. This effect becomes even more pronounced in narrow waveguides. 

\begin{figure}
\includegraphics[width=8.2cm]{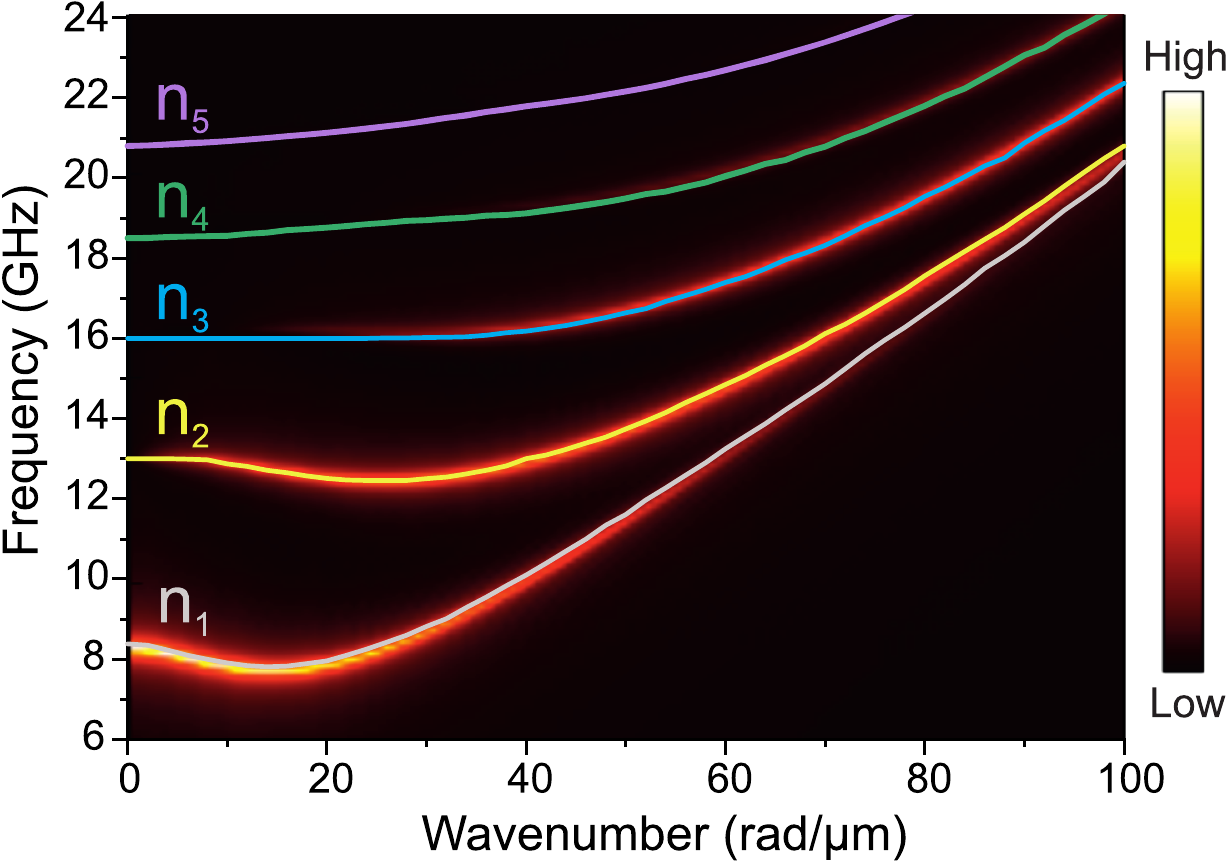}
\caption{Dispersion relations of quantized spin-wave modes in a 200-nm-wide CoFeB waveguide. The color map corresponds to the dispersion relations of spin waves excited by an inductive antenna whereas the solid lines correspond to the dispersion relations of spin waves excited by the magnetoelastic shear strain state. The external transverse magnetic bias field was $\mu_0 H_\mathrm{ext} = 50$ mT.}
\label{fig:dispersion}
\end{figure}

The spin-wave dispersion relations for the quantised width modes were extracted from the micromagnetic simulations by applying a rectangular excitation pulse of 20 ps duration and performing subsequently both a spatial Fourier transform of the magnetization dynamics in the $x$-direction and a Fourier transform in time. The results are presented in Fig.~\ref{fig:dispersion}, which indicates the mode-dependent spin-wave frequencies $f_n$ as a function of their wavenumber $k$. The solid lines correspond to spin waves excited by a magnetoelastic shear strain pulse, whereas the color plot relates to spin wave generation by an Oersted field pulse from an inductive antenna. As Fig.~\ref{fig:dispersion} indicates, both excitation schemes resulted in exactly the same dispersion relations for all quantised modes. This demonstrates that the spin-wave mode formation in the waveguide is independent of the excitation mechanism.  

\begin{figure}
\centering
\includegraphics[width=8.2cm]{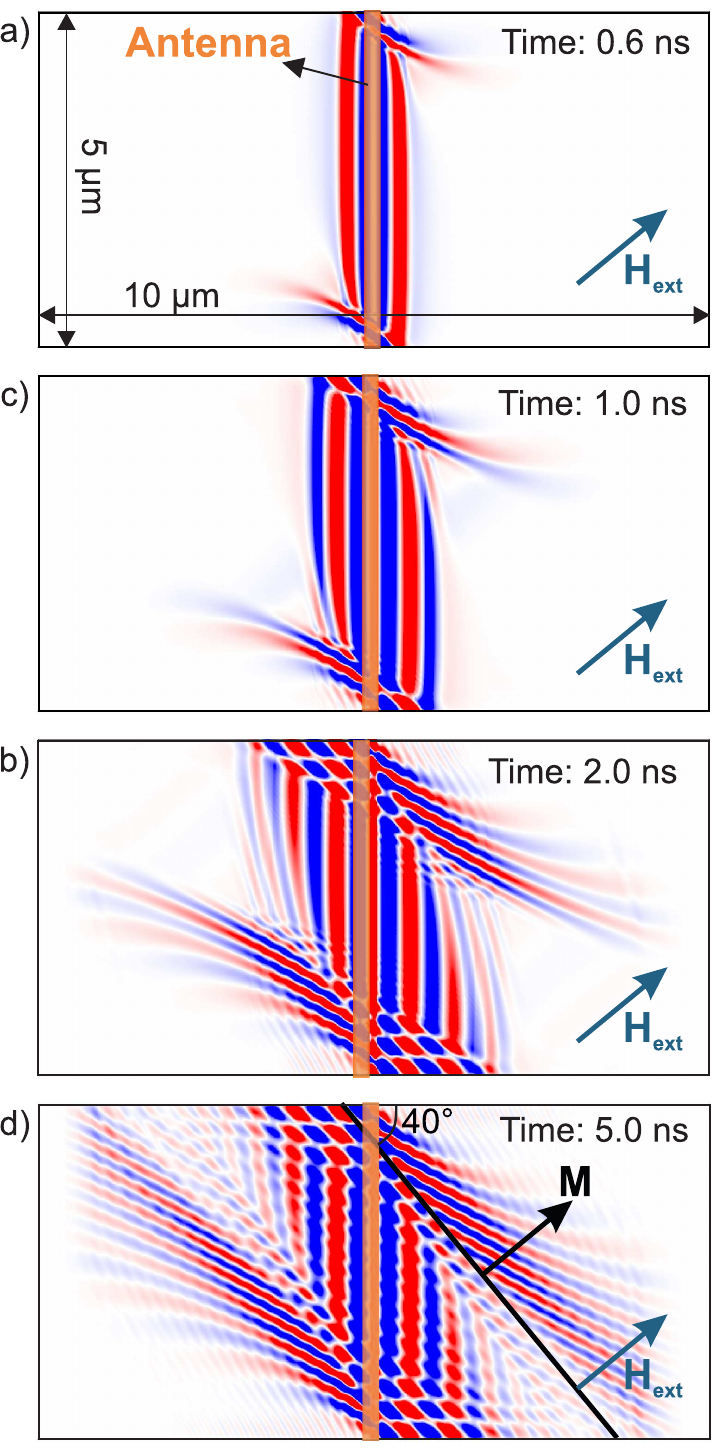}
\caption{\textbf{(a-d)} Snapshot images of the $z$-component of the magnetization dynamics excited by an inductive antenna in a 5-$\mu$m-wide CoFeB waveguide with an oblique bias field at an angle of $\theta = 40$\deg{}. The time is stated relative to the beginning of the excitation at $t = 0$. The excitation frequency was 13 GHz and the bias field was $\mu_0 H_\mathrm{ext} = 100$ mT.}
\label{fig:tilt}
\end{figure}

A key feature of the spin-wave modes in Figs.~\ref{fig:modes}a--c is the tilted wavefront, which means that phase and group velocities are not colinear. Since this is observed for both magnetoelectric and inductive transducers, it cannot be attributed to the excitation mechanism. Therefore, it must be rather a consequence of the magnetization orientation, which is not along one of the principal axes $\hat{x}$ or $\hat{y}$ of the waveguide. To shed light on this behavior, the spin-wave excitation in a 5-$\mu$m-wide CoFeB waveguide was studied with the magnetic bias field (and thus the magnetization) oriented under an angle $\theta$ with respect to the $x$-direction. The angle $\theta$ was set to the average angle of the equilibrium magnetization of the mesoscale 200-nm-wide waveguide (\emph{cf.} Fig.~\ref{fig:geo}), $\theta = \arctan(\Tilde{m}_y/\Tilde{m}_x)$ with $\Tilde{m}_y$ and $\Tilde{m}_x$ the magnetization components averaged over the waveguide width. For the waveguide geometry studied above, $\theta \approx 40$\deg{}. For simplicity, an inductive antenna was used in these simulations. Resulting magnetization snapshots at different times after the start of the excitation are displayed in Fig.~\ref{fig:tilt} for an excitation frequency of 13 GHz.

Figure \ref{fig:tilt} indicates that an Oersted field generated by an inductive antenna initially excites plane waves which gradually transform into a more complex pattern. For plane waves in an infinite film with uniform magnetization, the dynamic dipolar field has no component along the front of constant phase (here in the $y$-direction). For a waveguide, the situation is however different close to its edges. At the edges, as depicted in Fig.~\ref{fig:geo}d, the static magnetization becomes nonuniform. In addition, the boundary conditions also affect the \emph{dynamic} magnetization, rendering it also nonuniform. The resulting gradient of the dynamic magnetization in the $y$-direction leads to a nonzero transverse $y$-component of the dynamic dipolar field, $h^{(d)}_y$. In a thin film, for an edge at position $y = y_0$, $h^{(d)}_y$ can be written as

\begin{equation}
h^{(d)}_y = \frac{m^{(d)}_y}{\pi} \arctan\left( \frac{d/2}{y - y_0}\right).
\end{equation}

\noindent Here, $d$ is the thickness of the film and $m^{(d)}_y$ is the $y$-component of the dynamic magnetization.

As a consequence, this leads to isotropic secondary emission of spin wave at the edges, which then propagate towards the center of the waveguide. However, since the group velocity of the radiated spin waves is anisotropic, these secondary spin waves become strongly nonreciprocal (see Figs.~\ref{fig:tilt}b--d). The secondary spin waves then interfere with the plane wave leading to a tilt of the overall phase velocity with respect to the waveguide direction, as apparent in Fig.~\ref{fig:tilt}d.

For smaller waveguides, it is not possible anymore to distinguish clearly between edges and the center of the waveguide. Therefore, the secondary spin-wave emission is not spatially separated from the propagating plane wave. Together with the quantisation of the $y$-component of the wavevector, this leads to the formation of tilted spin-wave modes, as explained above. The tilt angle depends on the angular dependence of the group velocity of the spin waves and is in general not related to the (average) direction of the magnetization in a simple way. The detailed understanding of this relation is beyond the scope of this work and requires additional extensive micromagnetic simulations.

The propagation characteristics of spin waves in nanoscale waveguides with a transverse magnetic bias field and a tilted magnetization were discussed above for different excitation frequencies. Below, we address the relative excitation efficiency of different width modes. A striking feature is the observation of much stronger excitation of the second-order mode with respect to the first-order mode at higher frequencies. For example, this is indicated in Fig.~\ref{fig:modes}b for magnetoelastic excitation with the shear strain state at a frequency of 14 GHz. Note that the behavior was qualitatively independent of the excitation mechanism and the effect was also observed for magnetoelastic excitation with the normal strain state and inductive antennas (data not shown). This is in stark contrast to the conventional case of spin-wave excitation by an inductive antenna in a transversely (or longitudinally) magnetized waveguide where the magnetization is parallel to the external bias field. In these cases, the excitation of the second-order mode is forbidden by symmetry. In the following, we address the relative excitation efficiency of the individual spin-wave modes for the two different magnetoelastic strain states as well as inductive antennas as a reference.

The excitation efficiency of a specific spin-wave mode $n$, $A_n$, is proportional to the spatial overlap integral of the dynamic excitation field $\bm{H}_\mathrm{rf}(\bm{x})$ and the magnetization precession amplitude of mode $n$, $\bm{m}_n(\bm{x})$ \cite{demidov15}: 

\begin{equation}
\label{eq:eff_overlap}
A_n \propto \left| \int_V \bm{H}_\mathrm{rf}(\bm{x}) \cdot \bm{m}_n(\bm{x}) \: d\bm{x} \right|.
\end{equation}

For the case of an inductive antenna, the higher excitation efficiency of the second-order mode can be clearly linked to the tilting of the wavefront of the emitted spin waves. Since the magnetization profile of the first-order mode is (nearly) antisymmetric with respect to the waveguide width (\emph{cf.} Fig.~\ref{fig:modes}a), the overlap with the symmetric Oersted field from the inductive antenna---and thus according to Eq.~\eqref{eq:eff_overlap} the excitation efficiency---is (nearly) zero. In contrast, the magnetization profile of the second-order mode (\emph{cf.} Fig.~\ref{fig:modes}b) becomes nearly symmetrical due to the tilted wavefront and shows thus a large overlap integral with the exciting Oersted field. Hence, in a mesoscale waveguide, one finds the opposite of the case of a macroscopic waveguide with transverse field, namely that even modes are excited with high efficiency whereas odd modes are excited with low or zero efficiency.

The situation for magnetoelastic excitation fields is less intuitive since the exciting magnetoelastic field depends also on the magnetization direction (\emph{cf.}~Eq.~\eqref{eq:mel-field}). Figure~\ref{fig:geo}d indicates that the in-plane components of the magnetization in the mesoscale waveguide, $m_x$ and $m_y$, are symmetric. In contrast, the out-of-plane component, $m_z$, is antisymmetric. However, $m_z$ is rather weak and can thus be neglected. The strain is considered to be uniform over the waveguide width, as discussed earlier. Consequently, the magnetoelastic field is symmetric because the magnetization itself is also symmetric. Therefore, the qualitative behavior is similar to that of an inductive antenna resulting in a much larger excitation efficiency for even modes than for odd modes. 

In addition to its tilt with respect to the principal axes of the waveguide, the magnetization is also nonuniform. Hence, the magnetoelastic field becomes also nonuniform. Again, this is different from the case of an inductive antenna for which the exciting Oersted field is uniform and independent of the magnetization state of the waveguide. The nonuniformity of the magnetoelastic field is expected to modify the relative excitation efficiencies of the different modes.

\begin{figure*}
\centering
\includegraphics[width=15.8cm]{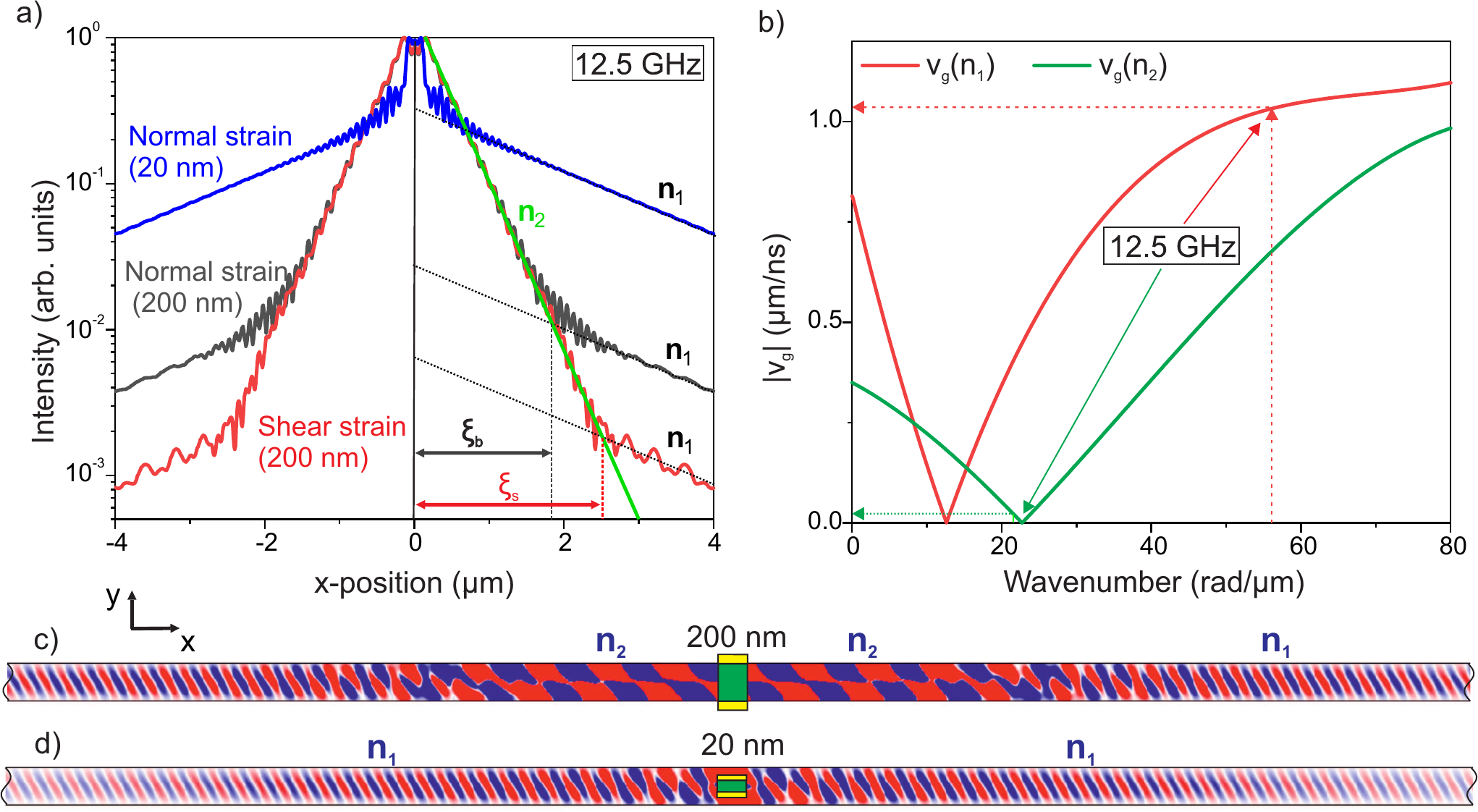}
\caption{\textbf{(a)} Spin-wave intensity along a 200-nm-wide CoFeB waveguide (in propagation direction) for different excitation fields and different excitation areas, as indicated. \textbf{(b)} Group velocity of the $n_1$ and $n_2$ spin-wave modes in the waveguide as a function of their wavenumber. \textbf{(c,d)} Steady-state snapshots of the $z$-component of the magnetization dynamics at 12.5 GHz for the magnetoelastic normal strain state with widths of 200 nm and 20 nm, respectively. The external transverse magnetic bias field was $\mu_0 H_\mathrm{ext} = 50$ mT.}
\label{fig:intensity}
\end{figure*}

To further assess the influence of the excitation field on the mode-dependent spin-wave excitation efficiency, we compare the magnetoelastic excitation fields based on the normal and shear strain states. For both cases, Fig.~\ref{fig:intensity}a indicates the calculated spin-wave intensity along the length of the waveguide averaged over the waveguide width at a frequency of 12.5 GHz. A steady-state magnetization snapshot for the normal strain state is depicted in Fig.~\ref{fig:intensity}c. For a better comparison, the excitation area was the same for the two strain configurations and the spin-wave intensity was normalized to the value in the excitation region.

Close to the excitation region, the spin-wave intensity contains contributions from several spin-wave modes. To quantify the relative intensities of the different modes, we take advantage of their different group velocities of the modes. The intensity of a single spin-wave mode is expected to decay exponentially during propagation along the waveguide. The intensity can then be described by

\begin{equation}
    \log \left( I_n(x) \right) = \log(I_{0n}) - \frac{x}{\delta_n},
\end{equation}

\noindent with $\delta_n$ the decay length of mode $n$. The spin-wave intensity along the waveguide in Fig.~\ref{fig:intensity}a indicates regions with different slopes, which can be attributed to different predominant propagating spin-wave modes with different decay lengths $\delta_n$. The mode-dependent decay length stems from the mode-dependent group velocity and is given by $\delta_n = v_{g,n} \tau$ with $v_{g,n} = 2\pi \partial f_n/\partial k$ the group velocity and $\tau$ the lifetime of the spin-wave mode. The lifetime is expected to be approximately the same for the two modes whereas the group velocity strongly differs. This is indicated in Fig.~\ref{fig:intensity}b as well as by the dispersion relation in Fig.~\ref{fig:dispersion}. At a frequency of 12.5 GHz, the first-order mode is much faster than the second-order mode and thus $\delta_1=4.64$~$\mu\mathrm{m} \gg \delta_2=0.88$~$\mu$m.

As graphically depicted in Fig.~\ref{fig:intensity}a, the second-order mode $n_2$ is excited with higher efficiency by magnetoelastic fields but decays faster with propagation distance than the first-order mode $n_1$. Hence, after a propagation distance of $\xi$, the intensity of two modes becomes equal. The value of $\xi$ depends on the decay lengths as well as the ratio of the initial intensities, $I_{0,1}$ and $I_{0,2}$, at the excitation region, \emph{i.e.} the relative excitation efficiencies, and is given by 

\begin{equation}
\label{intensity}
    \xi = \frac{\delta_1 \delta_2}{\delta_1- \delta_2} \log \left(\frac{I_{0,2}}{I_{0,1}} \right).
\end{equation}

\noindent Thus, $\xi$ can be used to quantify the relative excitation efficiencies of different spin-wave modes $I_{0,1}/I_{0,2}$. As discussed above, the excitation efficiency is much larger for the second-order mode and thus $I_{0,2} \gg I_{0,1}$. Figure~\ref{fig:intensity}a indicates that, at 12.5 GHz, $\xi$ was larger for shear strain than for normal strain state. Thus, the relative excitation efficiency of the second-order mode with respect to the first-order mode is larger for normal strain state.

\begin{figure}
\centering
\includegraphics[width=8cm]{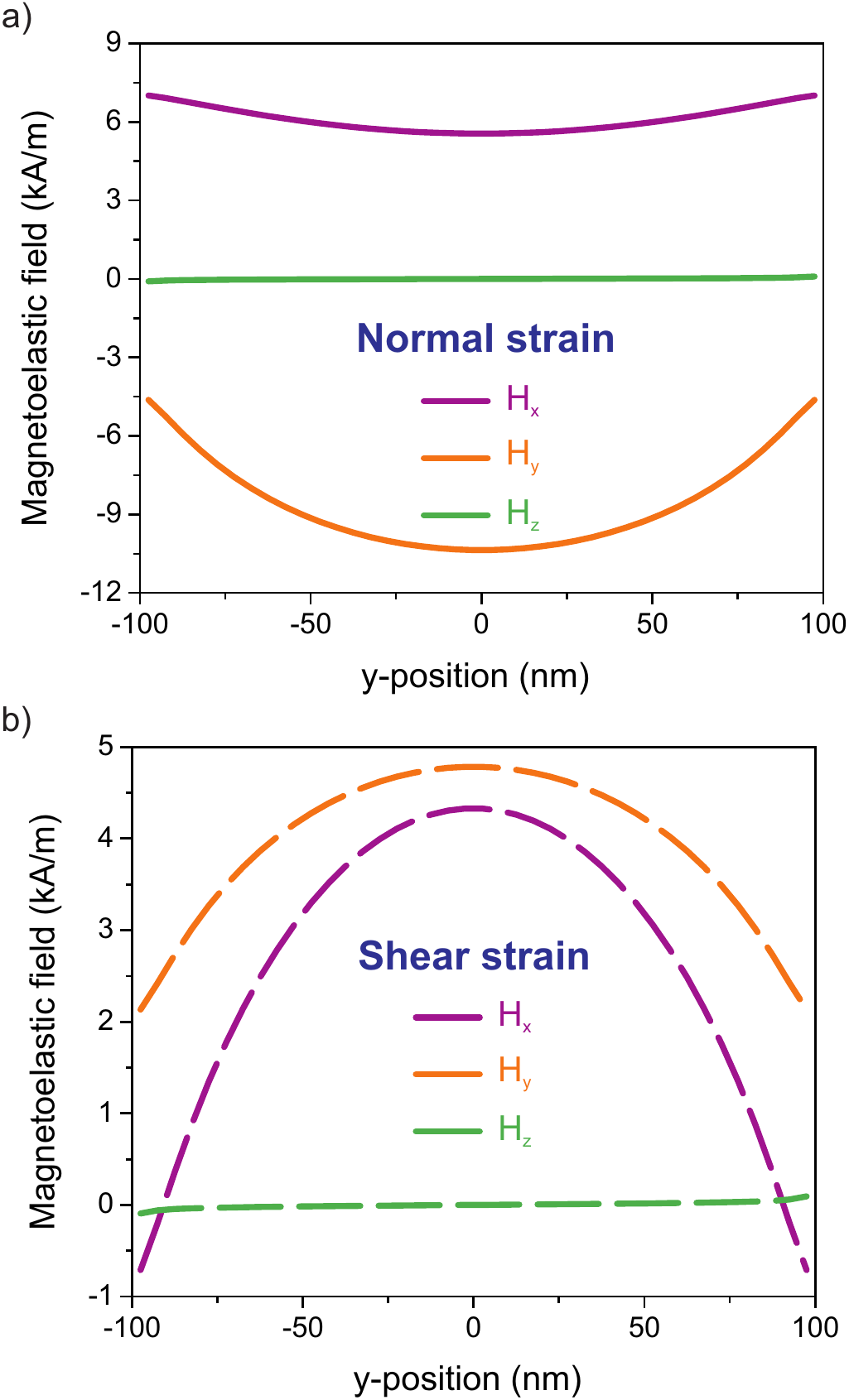}
\caption{Variation of the $x$-, $y$-, and $z$-components of the magnetoelastic field across a 200-nm-wide CoFeB waveguide generated by \textbf{(a)} normal and \textbf{(b)} shear strain state, respectively. The external transverse magnetic bias field was $\mu_0 H_\mathrm{ext} = 50$ mT.}
\label{fig:ME_field_profile}
\end{figure}

This behavior can be understood by considering that normal and shear strains couple to different components of the magnetization (\emph{cf.} Eq.~\eqref{eq:mel-field}). Figures~\ref{fig:ME_field_profile}a and \ref{fig:ME_field_profile}b indicate the different components of the magnetoelastic field as a function of the position across the waveguide for the normal and shear strain state, respectively. Since the magnetization in the waveguide is nonuniform across the width, the magnetoelastic fields are also nonuniform. Figure~\ref{fig:ME_field_profile}a indicates that the amplitude of the magnetoelastic field due to shear strain is maximal near the edges and minimal in the center of the waveguide. In contrast, the amplitude of the magnetoelastic field due to normal strains reaches a maximum in center and a minimum near the edges. Since the first- (second-) order spin-wave mode has highest (lowest) amplitude in the center and lowest (highest) near the edges of the waveguide, the overlap integral (\emph{cf.}~Eq.~\eqref{eq:eff_overlap}) with the magnetoelectric field due to normal strain is relatively larger (smaller) with respect to the magnetoelectric field due to shear strain. Thus, by selecting different components of the strain tensor in a magnetoelectric transducer, and therefore changing the geometry of the magnetoelastic field, it becomes possible to tune the relative excitation efficiency of different modes.

Equation~\eqref{eq:eff_overlap} indicates that the relative excitation efficiency of different spin-wave modes depends on the spatial overlap between the excitation field and the mode profiles. Therefore, changing the area and/or the geometry of the transducer can also be used to tailor the relative excitation efficiency or to preferentially excite a specific mode. To illustrate this, we have performed additional simulations using a much smaller excitation region with an area of 50 x 20 nm$^2$ (see Fig.~\ref{fig:intensity}d). The excitation frequency was 12.5 GHz as in the discussion above. The intensity along the length of the waveguide is depicted in Fig.~\ref{fig:intensity}a and indicates that the smaller area excites the first-order mode with much higher relative efficiency, leading to a much shorter $\xi$ (\emph{cf.} Fig.~\ref{fig:intensity}a). Since the second-order mode has a small amplitude in the middle of the waveguide, the overlap integral between the excitation field and the mode profile of $n_2$ is nearly zero. This demonstrates that mode-selective spin-wave excitation can be realized simply by adjusting the area of the magnetoelectric transducer.

In the previously explained, the relative excitation efficiency of different modes was discussed as a function of the strain geometry, shear \emph{vs.} normal. Here, we comment on the absolute excitation efficiencies of spin waves by different excitation fields, both magnetoelastic as well as Oersted fields. The magnetization dynamics in the excitation area described within the Landau-Lifshitz-Gilbert equation are determined by the torque exerted by the applied microwave magnetic field

\begin{equation}
\bm{\tau}(y) = \mu_0 \bm{m}(y) \times \bm{H}_\mathrm{rf}(y).
\end{equation}

\noindent The total torque per applied field $\tau_\mathrm{tot}(y)/(\mu_0 H_\mathrm{tot}(y))$ can be used to compare the total excitation efficiency of different excitation mechanisms. The magnetoelastic fields and Oersted fields generated by antenna have in general a rather different dependence on the device geometry and therefore may possibly exert very different torques on the magnetization, even if their magnitude is identical. In the waveguides considered here, the magnetization $\bm{m}$ is both tilted with respect to the principal axes as well as nonuniform across the width of the waveguide. As a result, the torque also becomes nonuniform across the width, analogously to the nonuniformity of the magnetoelastic fields. Nonetheless, the magnitude of $\tau_\mathrm{tot}(y)$ can be semiquantitatively compared for different excitation geometries even without defining an accurate averaging procedure. 

\begin{figure}
\centering
\includegraphics[width=8cm]{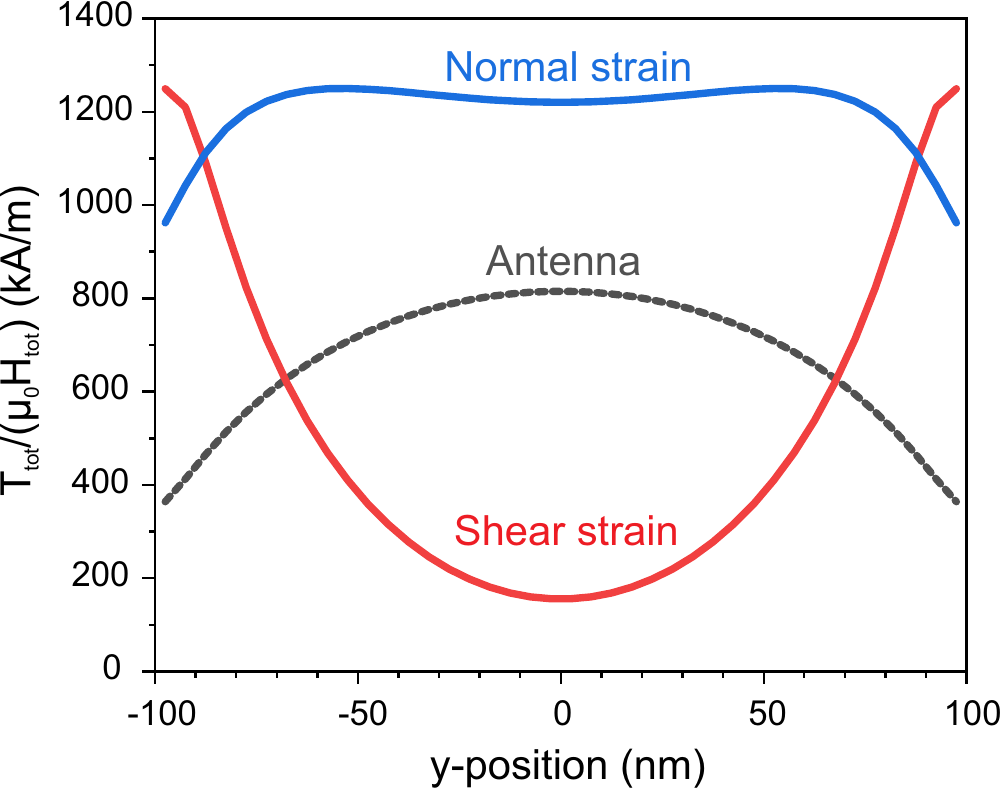}
\caption{Variation of the normalized torque $\tau_\mathrm{tot}/(\mu_0 H_\mathrm{tot})$ exerted on the magnetization in a 200-nn-wide CoFeB waveguide by an Oersted field due to an inductive antenna as well as by magnetoelastic fields due to normal and shear strain states, as indicated. The external transverse magnetic bias field was $\mu_0 H_\mathrm{ext} = 50$ mT.}
\label{fig:Torque_profile}
\end{figure}

Figure \ref{fig:Torque_profile} depicts the normalized torque $\tau_\mathrm{tot}(y)/(\mu_0 H_\mathrm{tot}(y))$ for magnetoelastic fields due to shear and normal strain states as well as for an Oersted field generated by an inductive antenna as a function of the position across the waveguide. The torque is larger for a magnetoelastic field due to normal strains than for an Oersted field at all positions of the waveguide. In agreement with the above discussion, magnetoelastic fields due to shear strain generated large torques at the edges of the waveguide. As a whole, we find, in the specific geometry considered here, that in-plane normal strain is most efficient to generate spin waves, whereas the magnitude of the torques due to shear strain and Oersted fields are similar. It is clear that the efficiencies due to the different mechanisms depend strongly on the tilt of the magnetization and therefore strongly on the waveguide width (the shape anisotropy field) and the applied bias field. As a simple example, the normalized torque generated by an Oersted field $\tau_\mathrm{tot}^\mathrm{(Oe)}/\left(\mu_0 H_\mathrm{tot}^\mathrm{(Oe)}\right)$ depends on the magnetization angle $\theta$ approximately as $\sin\theta$. A comprehensive discussion of the dependence of the spin-wave excitation efficiency is beyond the scope of this paper and will be the subject of a forthcoming study. The results however suggest that the nonuniformity and complex nature of the magnetoelastic fields is not necessarily detrimental to excite spin waves with respect to more conventional Oersted fields. Furthermore, the results indicate that the optimization of magnetization direction and strain geometry may be used to maximize magnetoelastic torques in mesoscale magnetic waveguides.

\begin{figure*}
\centering
\includegraphics[width=16.2cm]{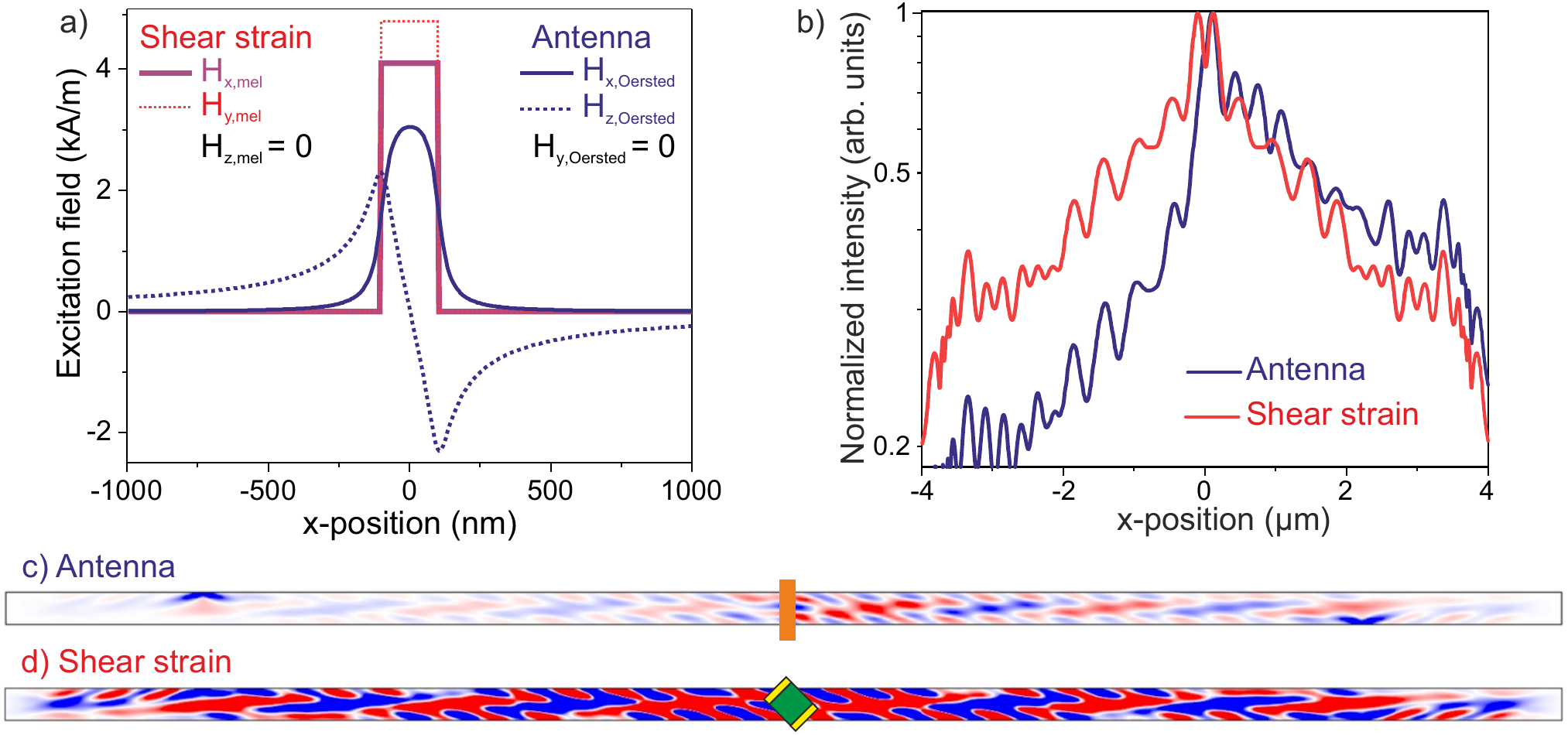}
\caption{\textbf{(a)} Oersted excitation field due to an inductive antenna as well as the  magnetoelastic excitation field due to shear strain extracted at the center of the magnetic waveguide. \textbf{(b)} Spin-wave intensity in a 200-nm-wide CoFeB waveguide along the propagation direction averaged over the waveguide for the excitation fields given by sub-figure \textbf{(a)}  at a frequency of 11 GHz. \textbf{(b,c)} Snapshot images of the $z$-component of the magnetization dynamics under steady state conditions for Oersted and magnetoelastic excitation. The external transverse magnetic bias field was $\mu_0 H_\mathrm{ext} = 50$ mT.}
\label{fig:non-reci}
\end{figure*}

We finally comment on the nonreciprocity of the generated spin waves by the different excitation mechanisms. It is well known that the nonreciprocal propagation of magnetostatic surface waves is enhanced by the chirality of the Oersted field generated by an inductive wire antenna \cite{demidov09}. In contrast, magnetoelastic fields are not expected to show chirality due to their symmetry, see Eq.~\eqref{eq:mel-field} and Fig.~\ref{fig:non-reci}a which displays both the Oersted and the magnetoelastic excitation field over the length of the waveguide. Hence, spin waves excited by magnetoelastic fields are not expected to be nonreciprocal, but to exhibit symmetric radiation patterns in the waveguide.

The behavior is illustrated in Fig.~\ref{fig:non-reci}b, which indicates the normalized intensity of spin waves excited by an Oersted field due to an inductive wire antenna and a magnetoelastic field due to shear strain. In these simulations, a larger waveguide thickness of 40 nm was used to enhance the nonreciprocity of the spin waves excited by the antenna. The variation of the spin-wave intensity for antenna excitation indicates clear signs of nonreciprocity with spin waves with much higher intensity traveling in the positive direction of the waveguide. In contrast, the intensity profile of spin waves excited by the magnetoelastic field is symmetric with respect to the transducer position at $x = 0$. We note that the mode profile is identical for the two excitation mechanisms (see Fig.~\ref{fig:non-reci}b and c for  steady-state magnetization snapshots).

In conclusion, we have analyzed the excitation of spin waves by the magnetoelastic effect via applying local normal or shear in-plane strains in mesoscopic waveguides. Since the shape anisotropy field in the waveguide was comparable to the external bias field, the static magnetization was tilted with respect to the principal axes of the waveguide and nonuniform over the width. The propagation characteristics of spin waves in such waveguides and their dispersion relation were significantly altered with respect to ideal Damon-Eshbach or backward-volume geometries. Independently of the excitation mechanism, quantised width modes with both odd and even mode numbers were observed. Moreover, all quantised spin-wave modes showed a tilted phase front with noncolinear phase and group velocities. Simulations of transient magnetization dynamics during the initial stages of spin-wave excitation indicated that the tilted phase front originated from the orientation of the average magnetization with respect to the principal axes of the magnetic waveguide. The tilt of the magnetization led to additional excitation of secondary spin waves at the edges of the waveguide; the interference with the plane wave initially generated in the center of the waveguide led to the tilted phase front observed in later steady-state conditions.

In addition, the excitation efficiency of the second-order mode was found to be generally larger than that of the first-order mode. This was linked to the tilt of the phase front, which affects the mode profile and consequently also the excitation efficiency, as expressed by the the overlap integral with the excitation field. The relative intensity of the excited modes could be controlled by the strain state as well as by tuning the dimensions of the excitation area. These characteristics were further compared to the more conventional excitation of spin waves via an Oersted field emitted by a wire antenna. We indicate that the two excitation mechanisms generate spin waves with similar absolute efficiency despite their different geometries. Moreover, the nonreciprocity of the spin-wave excitation due to the chirality of the Oersted field was removed by using a magnetoelastic excitation.

\begin{acknowledgments}

This work has been supported by imec’s industrial affiliate program on beyond-CMOS logic. It has also received funding from the European Union’s Horizon 2020 research and innovation program within the FET-OPEN project CHIRON under grant agreement No. 801055. F.V. acknowledges financial support from the Research Foundation –- Flanders (FWO) through grant No. 1S05719N.
\end{acknowledgments}

\clearpage

\end{document}